\documentclass[12pt]{article}
\usepackage{a4wide, amsmath, latexsym, epsfig,amsfonts,
  psfrag,graphicx, rotating, fancyhdr, tabularx, afterpage,booktabs}
\usepackage{subfigure}
\usepackage{amssymb}
\usepackage{setspace}
\newcommand\nn{\nonumber}
\newcommand\ba{\begin{eqnarray}}
\newcommand\ea{\end{eqnarray}}

\begin{document}
\begin{titlepage}
 
\begin{flushright} 
{ IFJPAN-IV-2012-13 \\ CERN-PH-TH-2012-347} 
\end{flushright}

\vspace{0.2cm} 
\begin{center}

{\Huge \bf Ascertaining the spin for new resonances decaying into ${\mathbb\tau^+ \tau^-}$  at Hadron Colliders}
\end{center}
\vspace*{5mm}

\begin{center}
   {\bf S. Banerjee$^{a}$,  J. Kalinowski$^{b}$,  W. Kotlarski$^{b}$,  T. Przedzinski$^{c}$  and Z. W\c{a}s$^{d,e}$}\\
{\em $^a$ Department of Physics, University of Wisconsin, Madison WI  53706, USA.}\\
{\em $^b$ Faculty of Physics, University of Warsaw, Warsaw, ul.\ Ho\.za 69, Poland.}\\
{\em $^c $ Faculty of Physics, Astronomy and Applied Computer Science,\\
Jagellonian University, Reymonta 4, 30-059, Krak\'ow, Poland.}\\
{\em $^d$ Institute of Nuclear Physics, PAN, Krak\'ow, ul. Radzikowskiego 152, Poland}\\
{\em $^e$ CERN PH-TH, CH-1211 Geneva 23, Switzerland.}
\end{center}
\vspace{.1 cm}
\begin{center}
{\bf   ABSTRACT  }
\end{center} 

Evidence of a new particle  with mass $\sim$125 GeV decaying into a pair of tau leptons at the Large Hadron Collider spurs interest in ascertaining its spin in this channel.
Here we present a comparative study between spin-0 and spin-2 nature of this new particle, using spin correlations and decay product directions.
The {\tt TauSpinner} algorithm is used to re-weight distributions from $q \bar q \to \gamma/Z \to \tau^+ \tau^-$ sample to simulate a spin-2 state exchange.
The method is based on supplementing the Standard Model matrix elements with those arising from presence of a new interaction.
Studies with simulated samples demonstrate the discrimination power between these spin hypotheses based on data collected at the Large Hadron Collider.
 
 \vspace{1cm}
\begin{flushleft}
{   IFJPAN-IV-2012-13 \\  CERN-PH-TH-2012-347\\
 December, 2012}
\end{flushleft}
 
%%%%%%%%%%%%%%%%%%%%%%%%%%%%%%%%%%%%%%%%%%%%%%%%%%%%%%
\vspace*{1mm}
\bigskip
%%%%\vfill
\footnoterule
\noindent
{\footnotesize \noindent% 
Email addresses: Swagato.Banerjee@cern.ch, Jan.Kalinowski@fuw.edu.pl, Wojciech.Kotlarski@fuw.edu.pl, Tomasz.Przedzinski@cern.ch and Z.Was@cern.ch.
}
\end{titlepage}

\section {Introduction}

Following the discovery of a Higgs-like state $H$ with mass $\sim$~125 GeV at the Large Hadron Collider (LHC)~\cite{ATLAS_disc,CMS_disc},
its spin-parity assignment must be  examined to verify the true nature of this new state. 
The spin of this newly observed state has recently been discussed~\cite{VB} in the context of its couplings to  a pair of vector bosons.
However, from an experimental point of view, the spin property should be investigated channel by channel, and other alternative hypotheses should be investigated and excluded.  
At the HCP'12 conference, the ATLAS~\cite{ATLAS_HCP} and the CMS~\cite{CMS_HCP} collaborations reported observed significances of 1.1 $\sigma$ and 1.5 $\sigma$ respectively,
for the  $H\to\tau^+\tau^-$  decay channel. Their corresponding expected significances are 1.7 $\sigma$ and 2.5 $\sigma$, 
which when added in quadrature are already at the 3 $\sigma$ level.
In the present paper, we concentrate on this channel as a possible completion of the spin studies.  

Searches for $H\to\tau^+\tau^-$ decay are challenging because the $\tau$ neutrino's escape detection.
Experimental signatures are categorized over multiple channels in terms of  observable final state decay products.
Data from the multi-channel inputs must be compared with simulation of large samples of Monte Carlo (MC) events, 
which includes detector resolution and acceptance effects, as well as contributions from background events in the selected sample.

The study of $\tau$ polarization can provide additional leverage for this search.
The  {\tt TauSpinner} algorithm~\cite{Czyczula:2012ny} provides a mechanism to evaluate the polarization effects of $\tau$ spin.
The algorithm based on re-weighting technique can be applied to existing sample of simulated MC events,
thereby reducing the need for computationally intensive simulation of independent samples,
and has successfully been applied for measurements of $\tau$ polarization in $W^\pm\to\tau^\pm\nu$~\cite{:2012cu}  and $Z\to\tau^+\tau^-$~\cite{Deigaard} decays.

In present paper we  extend the method of {\tt TauSpinner} by adding contributions 
from new resonances to the amplitude of $q \bar q \to \gamma/Z  \to \tau^+\tau^-$ processes. 
Our numerical study  based on exchange of a spin-2 state X is motivated by recent interest 
in measurement of spin properties for the newly discovered Higgs-like particle candidate.
In general, contribution from other new interactions, such as those arising from an additional $Z'$ boson, can also be evaluated in this way. 
Though the present implementation illustrates re-weighting of samples generated with Pythia~\cite{Sjostrand:2007gs}, 
the method is equally applicable to other MC event generators.

Our paper is organized as follows. In the next section we discuss contributions from spin-2 state to the matrix element.
Section 3 is devoted to the {\tt TauSpinner} algorithm and the inclusion of new matrix elements into the program.
Section 4 is devoted to technical tests of the {\tt TauSpinner} algorithm and stability tests of internal cross-checks.
In section 5, we investigate experimentally discriminating variables sensitive to spin.
In section 6, we perform a numerical analysis to access the sensitivity to measure the spin properties of Higgs-like states. 
Section 7 presents the Summary and Appendix A closes the paper with detailed description of updates to the  {\tt TauSpinner} algorithm.
 
\section {Quark level cross section for $\gamma/Z/X$ production of tau pairs. } \label{noSM}

In many theoretical models massive objects of spin-2 arise, including KK gravitons~\cite{ADD,RS}, 
analogues of the $f_2$ state of QCD in a new strongly-interacting sector~\cite{MP} or states in four-dimensional ghost-free models of massive gravity~\cite{RG}.
For our purposes we will treat the spin-2 particle with mass of 125 GeV as a low-energy signature of some unspecified high-energy completion of the model. 
Therefore, we will use an effective Lagrangian formalism  for a spin-2 field interacting with fermions 
to calculate the angular distribution in the process $q\bar q\to \gamma/Z/X\to\tau^+\tau^-$ at the lowest level.  
These quark level calculations are implemented in the {\tt TauSpinner}~\cite{Czyczula:2012ny} algorithm, as described in Section 3.

 For a symmetric spin-2 field $X_{\mu\nu}$ with mass $M_X$, decay width $\Gamma_X$  and momentum $k$ the propagator reads~\cite{vDV} as:
 \begin{equation}
 \Delta^{\mu\nu,\alpha\beta}(k)=\frac{iP^{\mu\nu,\alpha\beta}}{k^2-M^2_X+iM_X\Gamma_X}.
 \end{equation}
The  projector $P$ is given by:
 \begin{equation}
 P^{\mu\nu,\alpha\beta}=\frac{1}{2}(\eta^{\mu\alpha}\eta^{\nu\beta}+\eta^{\mu\beta}\eta^{\nu\alpha})- \frac{1}{3}\eta^{\mu\nu}\eta^{\alpha\beta} +\ldots,
 \end{equation}
where $\eta^{\mu\nu}$ is the Minkowski tensor.  
The terms proportional to the momentum $k$, represented by dots in the above formula,  will vanish when contracted with the on-shell massless fermion currents.

The interactions of $X^{\mu\nu}$ with fermions consists of various operators of increasing dimensions, 
suppressed by powers of some high scale denoted by $F$.  At zero derivative level the coupling of $X$ to a fermion current has a form:
\ba
{\cal L} \ni  X^\mu_\mu \bar \psi( \lambda_L P_L +\lambda_R P_R) \psi +h.c., \label{eq:scal}
\ea  
where $P_{L,R}=(1\mp\gamma_5)/2$. 
This form of coupling is similar to an ordinary Yukawa coupling of a Standard Model (SM) singlet scalar.  
Therefore its experimental signatures: angular distributions and spin correlations,  
will be similar to a  scalar exchange, with the only difference coming from the spin-2 propagator.  
If the $X^{\mu\nu}$ were of gravity or extra dimension  origin, 
the couplings $\lambda_i$ for light fermions would naturally be suppressed by the fermion mass, $\lambda\sim m/F$.    
Therefore, in the discussion of the $\tau$-lepton pair production via Drell-Yan process in $pp$ collisions such couplings will be ignored.  

At dimension 5 level the coupling of $X^{\mu\nu}$ to a fermion bilinear is given by~\cite{fermioncoup,GMPU}:
\ba
{\cal L} \ni \frac{i}{4}\,\frac{1}{F} X^{\mu\nu}[\alpha^L\bar \psi_L (\gamma_\mu\partial_\nu+\gamma_\nu\partial_\mu)  \psi_L +\beta^L(\partial_\nu\bar \psi_L \gamma_\mu+\partial_\mu\bar \psi \gamma_\nu) \psi_L ]+(L\to R) +h.c.
\ea
The other possible dimension 5 coupling $\sim X^\mu_\mu \bar\psi \partial\!\!\!\slash \psi$ will be ignored 
since  for the on-shell fermions can be reduced to  the form in eq.(\ref{eq:scal}).

The couplings $\alpha^{L,R},\beta^{L,R}$ are model dependent.  
Although we do not attempt to construct any specific model, we assume that the couplings are quark- and lepton-flavor diagonal and,   
following Ref.~\cite{GMPU}, we make a simplifying assumption:
\ba
\alpha^{L,R}=-\beta^{L,R}\equiv g^{L,R}.
\ea 

At tree level the process $q\bar q\to \tau^+\tau^-$ proceeds via s-channel $\gamma/Z/X$ exchanges. 
The angular distribution in the CM frame can be written as a sum of the SM contribution:
\begin{eqnarray}
\frac{{\rm d}\sigma^{SM}}{{\rm d}\cos\theta}&= 
% gamma^2 
& \frac{e^4}{384 \pi  s} 
\left[\left({g^L_{\gamma qq}}^2 {g^L_{\gamma \tau \tau}}^2+
{g^R_{\gamma qq}}^2 {g^R_{\gamma \tau \tau}}^2\right)
(1+\cos\theta )^2\right. \nn \\
&&\phantom{M}+\left.
\left({g^L_{\gamma qq}}^2 {g^R_{\gamma \tau \tau}}^2+
{g^R_{\gamma qq}}^2 {g^L_{\gamma \tau \tau}}^2\right)
(1-\cos\theta)^2
\right]\nn \\
%gamma-Z  
  && -\frac{e^2}{192\pi}\, \frac{
(M^2_Z-s)} { 
(M_Z^2-s)^2+M_Z^2\Gamma_Z^2}  \nn \\
&& 
\phantom{M}\left[\left({g^L_{\gamma qq}}^{\phantom{1}}
g^L_{Zqq} g^L_{\gamma \tau \tau} g^L_{Z\tau\tau}
+g^R_{\gamma qq}g^R_{Zqq} g^R_{\gamma \tau \tau} g^R_{Z\tau\tau}\right)
(1+\cos\theta)^2 \right. \nn  \\
&& \phantom{Ma}
 \left.  +\left({g^L_{\gamma qq}}^{\phantom{1}}g^L_{Zqq} g^R_{\gamma \tau \tau} g^R_{Z\tau\tau}
+g^R_{\gamma qq}g^R_{Zqq} g^L_{\gamma \tau \tau} g^L_{Z\tau\tau}\right)
  (\cos\theta-1)^2\right] \nn \\
% Z^2
&& +\frac{1}{384\pi}\, \frac{s}{ 
(M_Z^2-s)^2+M_Z^2\Gamma_Z^2}\nn \\
  &&
\phantom{M}\left[\left({g^L_{Zqq}}^2 {g^L_{Z\tau\tau}}^2 +{g^R_{Zqq}}^2 {g^R_{Z\tau\tau}}^2\right)
  (1+\cos\theta)^2 \right. \nn \\
&& \phantom{Ma} \left.  +\left({g^L_{Zqq}}^2 {g^R_{Z\tau\tau}}^2 
+{g^R_{Zqq}}^2 {g^L_{Z\tau\tau}}^2\right) (\cos\theta-1)^2 \right]\label{eq:SM}
\end{eqnarray}
and a new term from the $X$ particle exchange, including its interference with the SM contribution. 
For a real field $X$ and real couplings $g^i_{Xff}$ it reads:
\begin{eqnarray}
\frac{{\rm d}\sigma^{X}}{{\rm d}\cos\theta}&= 
%  X^2
& \frac{1} {24576
\pi  f^4}\, \frac{s^3}{(M_X^2-s){}^2+M_X^2 \Gamma _X^2}\nn \\
&&
\phantom{M} \left[\left({g^L_{Xqq}}^2 {g^L_{X\tau \tau}}^2+{g^R_{Xqq}}^2 {g^R_{X\tau \tau}}^2\right)
(-1+2 \cos^2\theta+\cos\theta)^2 \right. \nn \\
&& \phantom{Ma}\left.
+\left({g^L_{Xqq}}^2 {g^R_{X\tau \tau}}^2+{g^R_{Xqq}}^2 {g^L_{X\tau \tau}}^2\right)
(1-2 \cos^2\theta+\cos\theta)^2\right]\nn \\
% gamma-X
&&   -\frac{e^2}{1536\pi f^2}\, \frac{ s
  (M_X^2-s)}{
  (M_X^2-s)^2+M_X^2 \Gamma _X^2}\nn \\
 && 
 \phantom{M} \left[\left({g^L_{Xqq}}^{\phantom{1}} g^L_{\gamma qq}g^L_{X\tau\tau} g^L_{\gamma \tau \tau }
  +g^R_{Xqq} g^R_{\gamma qq}g^R_{X\tau\tau} g^R_{\gamma \tau \tau}\right)
(1+\cos\theta)^2 (2 \cos\theta-1) \right. \nn \\
&& \phantom{Ma} \left. 
+\left({g^L_{Xqq}}^{\phantom{1}} g^L_{\gamma qq}g^R_{X\tau\tau} g^R_{\gamma \tau \tau}
  +g^R_{Xqq} g^R_{\gamma qq}g^L_{X\tau\tau} g^L_{\gamma \tau \tau}\right)
  (1+2 \cos\theta) (\cos\theta-1)^2\right]\nn \\
% Z-X
&&  +\frac{1}{1536\pi f^2}\,\frac{s^2 ((M^2_Z-s)(M_X^2-s)+M_Z\Gamma_Z M_X \Gamma _X) }
  {((M_Z^2-s)^2 
  +M_Z^2 \Gamma_Z^2)
((M_X^2-s){}^2+M_X^2 \Gamma _X^2)}\nn \\
&&\phantom{M}
\left[\left({g^L_{Xqq}}^{\phantom{1}} g^L_{Zqq}g^L_{X\tau\tau} g^L_{Z\tau\tau}+
g^R_{Xqq} g^R_{Zqq}g^R_{X\tau\tau} g^R_{Z\tau\tau}\right)
(1+\cos\theta)^2 (2 \cos\theta-1) \right. \nn \\
&&\phantom{Ma} \left.
+\left({g^L_{Xqq}}^{\phantom{1}} g^L_{Zqq}g^R_{X\tau\tau} g^R_{Z\tau\tau}+
g^R_{Xqq} g^R_{Zqq}g^L_{X\tau\tau} g^L_{Z\tau\tau}\right)
 (1+2 \cos\theta) (\cos\theta-1)^2\right].\label{eq:newX}
\end{eqnarray}
In the above expressions $\theta$ is the CM scattering angle of $\tau^-$ with respect to the incoming quark momentum, 
and $g_{Aff}^i$ (with $i=L,R$) denote the chiral couplings of $A=\gamma,Z,X$ bosons to a fermion  current $\bar f f$.  
In particular, for the SM couplings we have $g^i_{\gamma ff}=Q_f$, $g^i_{Zff}= \frac{g}{\cos\theta_W}(T_f^3-Q_f \sin^2\theta_W)$, 
with $g$ weak coupling constant, and $Q_f,\, T^3_f$ are the fermion electric charge and third component of weak isospin.  

The angular distribution due to the $X$ state exchange depends on its couplings. 
For general chiral couplings, the diagonal term is a fourth order polynomial in 
$\cos\theta$, and the interference  a third order polynomial. For vector or axial-vector 
type couplings ($g_{Xff}^L=\pm g_{Xff}^R$), the expressions in eq.(\ref{eq:newX}) simplify and the diagonal term 
exhibits a well known angular distribution $1-3\cos^2\theta+4\cos^4\theta$, while 
the interference with the SM contribution behaves as $\cos^3\theta$ 
and vanishes after angular integration. 
This can be observed in Fig.~\ref{Cstarprime_ZXH}  discussed later in the paper. For general couplings the interference survives the angular integration.

The user may take his/her preferred scenario and modify the parameters, including the SM ones, in these new currents.
For the sake of numerical comparisons, we use F = 1000 GeV to study the effects of the spin-2 state $X$ at 125 GeV.
The width of this state is taken to be 1.5 GeV.

Unless otherwise mentioned, for the default comparison, we set both the left-handed and right-handed couplings to have the  strengths of unity.
We also study two alternative scenarios of pure left-handed or pure right-handed couplings labeled in subsequent discussions as $L$ and $R$ respectively.
For these alternative models, the non-vanishing coupling constants are increased by $\sqrt{2}$ 
to allow direct comparison with the model $(L+R)$ where both couplings contribute equally.

\section{Algorithm case of  non-Standard Model weight}

Basic functionality of {\tt TauSpinner} relies on calculation of effective Born cross sections. It is documented in detail in Ref.~\cite{Czyczula:2012ny}. 
For the extension of {\tt TauSpinner} algorithm, starting with the case of production of $q \bar q \to \gamma/Z \to \tau^+\tau^-$ events,  
we replace the effective Born level contribution with the one where addition amplitudes due to the extra interaction are added. 
At present, the algorithm assumes that the resulting new cross-section has contributions 
from polynomials at most of the 4$^{th}$ order in $\mathrm cosine$ of the scattering angle $\theta$.
Spin correlations between the produced $\tau^+$ and $\tau^-$ remain as in the case of $\gamma/Z$ production, but differ only in angular dependence of the $\tau$ polarization.
Thus, the non-Standard Model weight can easily be calculated using the algorithm described in Appendix B.2 of~\cite{Czyczula:2012ny}.
With minor modifications as interfaced with the {\tt Tauola++} MC event generator~\cite{Davidson:2010rw}, the algorithm is described below:

\begin{itemize}
\item Initialize {\tt Tauola++}
\item Initialize {\tt TauSpinner}\\
      At this step, user provides two additional flags, {\tt nonSM2} and
      {\tt nonSMN}. The first one activates calculation of non-Standard Model weight. 
     Then effects due to spin correlations, described by the ratio  of {\tt WT1} and {\tt WT2}\footnote{{\tt WT1} is the spin weight of Standard Model
      and {\tt WT2} is when non-Standard Model interactions are switched on. {\tt WT1} is the default spin weight that {\tt TauSpinner} returns.},
      and effect on angular dependence in $\tau^+\tau^-$ production is introduced with {\tt WT3}.
      The factor due to the ratio of angle-integrated cross-section of SM and nonSM case is removed from {\tt WT3} if {\tt nonSMN} is set to 1. 
      All other aspects of {\tt TauSpinner} remain unchanged. 
\item Read the data files
\item Calculate the weights \\
      For the new non-Standard Model case, this step is extended:
      \begin{itemize}
        \item As in default case, calculate spin weight {\tt WT1} for Standard Model spin correlations
        \item Calculate spin weight {\tt WT2} for spin correlations calculated in non-Standard Model case.
              Use {\tt setNonSMkey(1)} to switch to non-Standard Model calculation mode.
        \item Calculate non-Standard Model weight {\tt WT3} for effects on cross section; 
                 use {\tt double getWtNonSM()} for its calculation.
        \item return {\tt WT = (WT2/WT1) * WT3}
      \end{itemize}
\end{itemize}

An example of such a calculation is given in our {\tt tau-reweight-text.cxx}.  
Weight calculation is prepared for the case when the generated sample has already the SM spin correlations taken into account.
User is free to use his own version of the function providing the Born level $\tau$ pair production and featuring another assumptions for SM extension. 
The necessary methods are described in Appendix~\ref{sec:TauSpinnerChanges} 

\section{Technical tests}

Before any use of our program is started, some technical tests must be performed
to verify if the function used in the implementation of non-Standard Model
interaction is proper from the point of view of our program requirements.
The first  is to check  the user provided  non-Standard Model Born cross-section. 
In particular, it has to be checked that the same conventions for input parameters as those used in {\tt TauSpinner} and  {\tt Tauola++} are chosen.
In our case, for the model as described in  previous sections, adjustment of signs for spin states had to be introduced.
The C++ function {\tt nonSM{\_}adopt} is prepared for the adjustment of conventions.

\begin{figure}[htp!]
\begin{tabular}{ccc}
  \includegraphics[width=0.48\columnwidth]{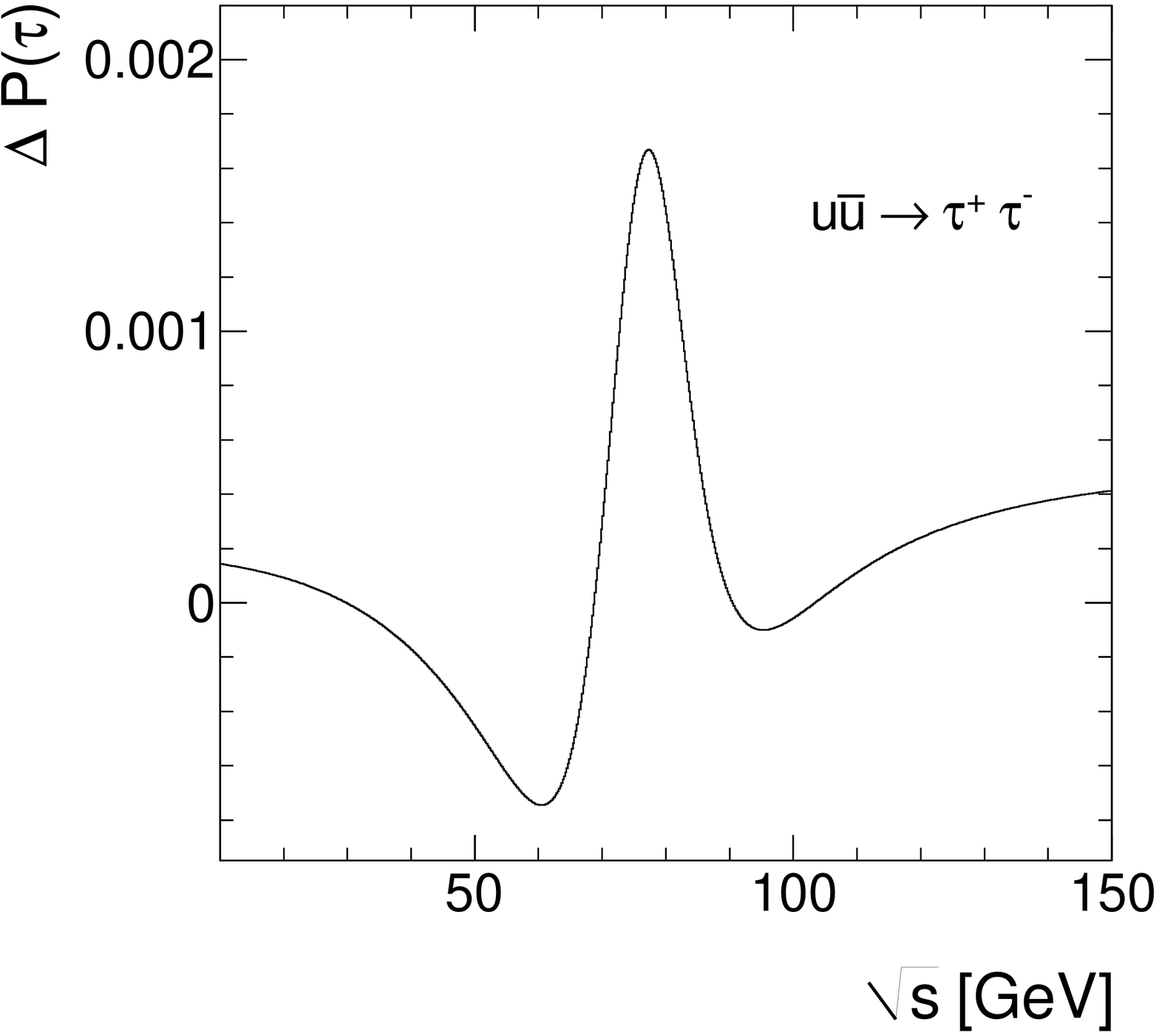} &
  \includegraphics[width=0.48\columnwidth]{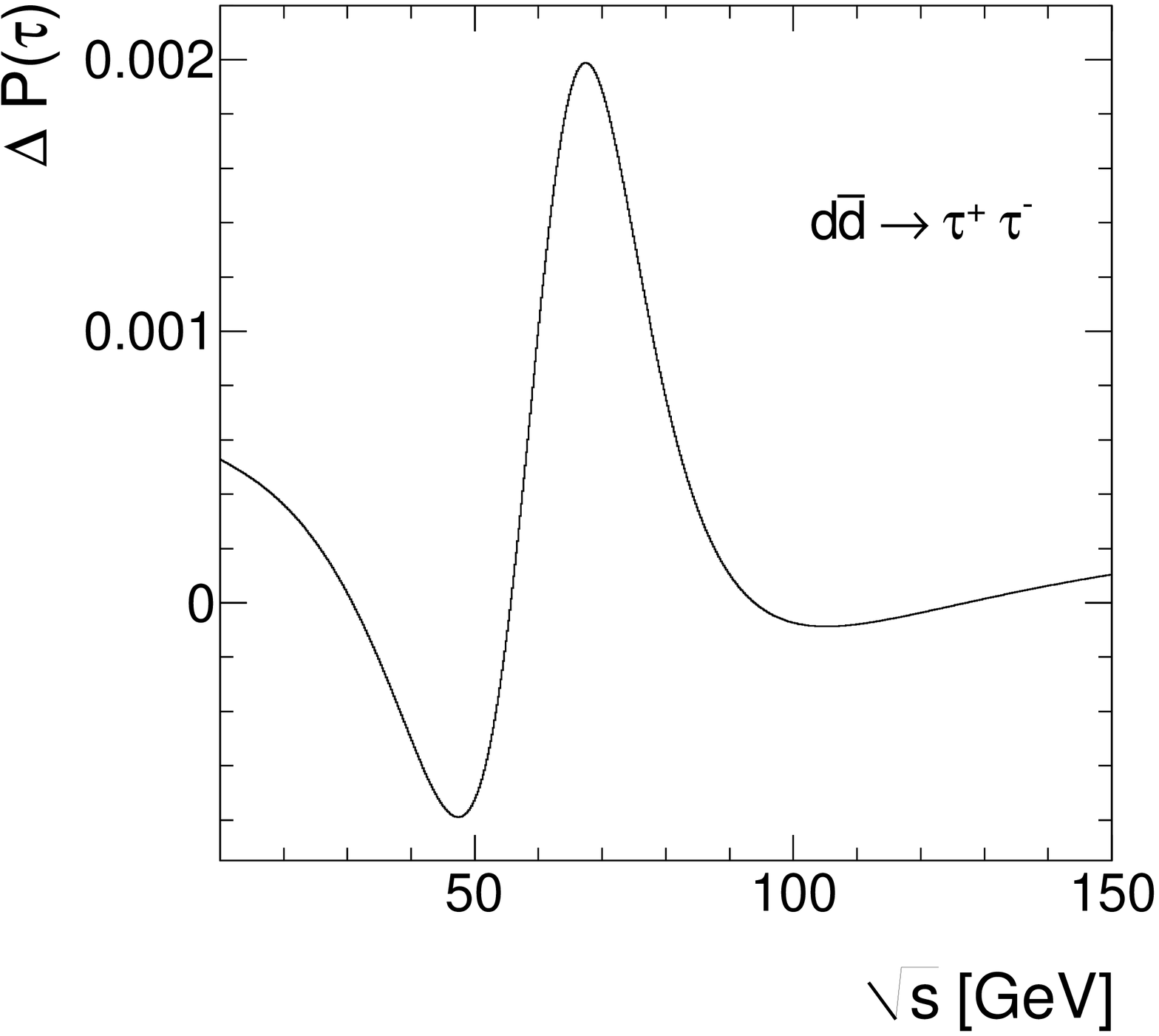} 
\end{tabular}
\caption{
The  difference between the $\tau$ polarization calculated  analytically 
from formulae as described in Section~\ref{noSM} but with new effects switched off, 
and the default implementation in the {\tt Tauola++} are plotted for 
fixed  scattering angle  $\cos\theta=0.3$ as a function of $\sqrt{s}$.
The plot on the left is for up quarks, and the plot on the right is for down quark. 
\label{figTest1}}
\end{figure}

\begin{figure}[htp!]
\begin{tabular}{ccc}
\includegraphics[width=0.48\columnwidth]{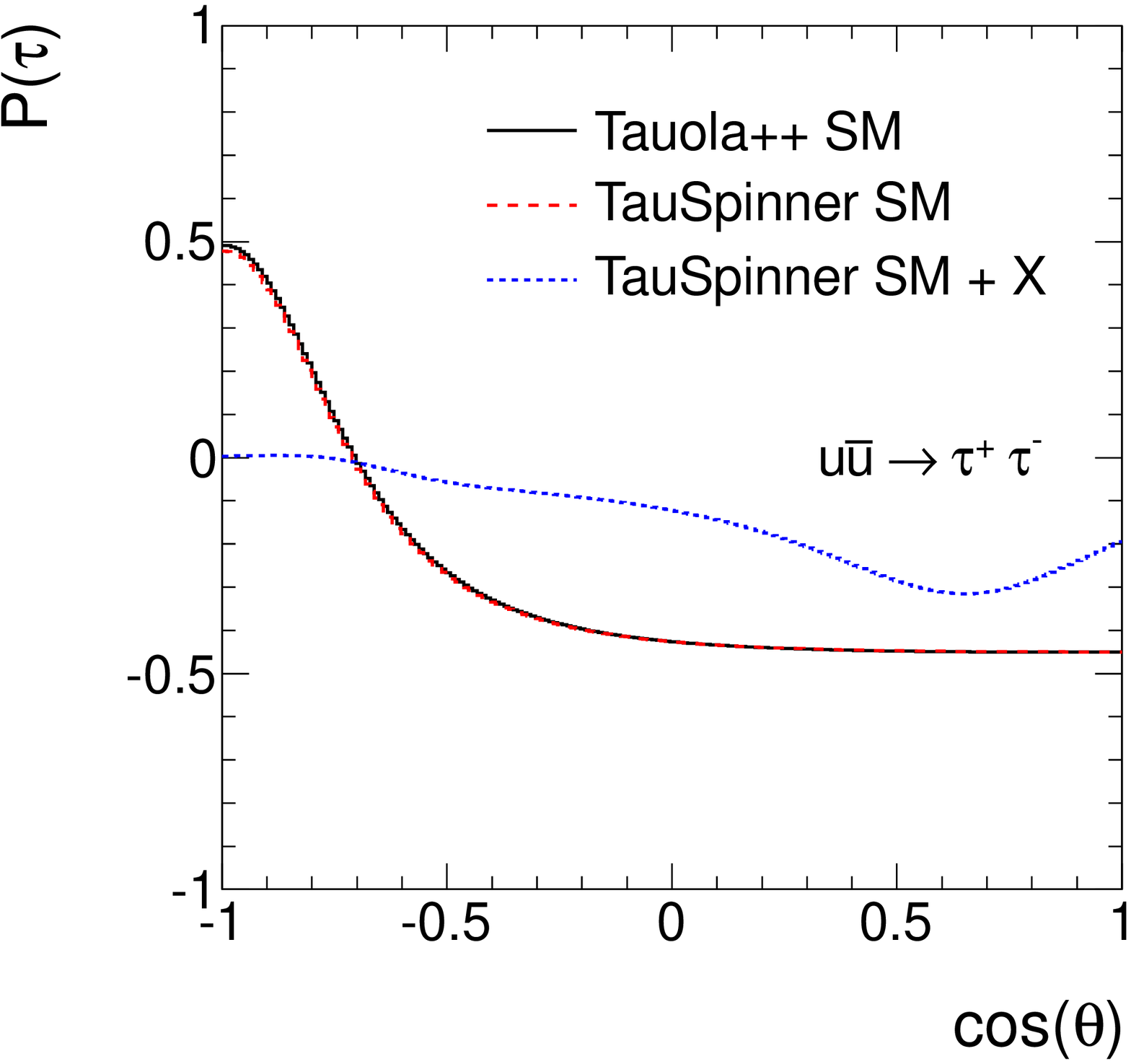} & 
\includegraphics[width=0.48\columnwidth]{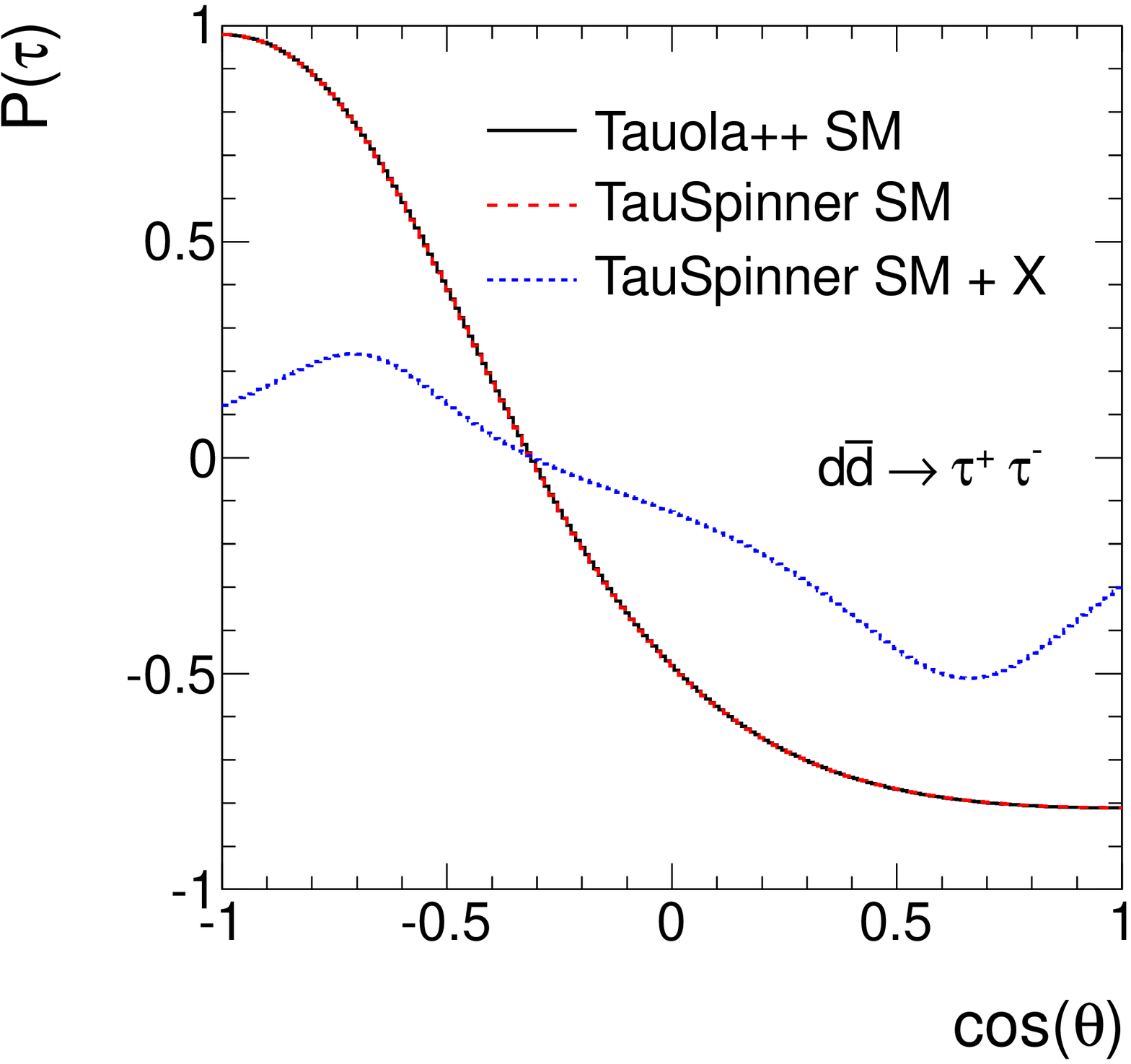} 
\end{tabular}
\caption{ Angular dependence of the $\tau$ polarization in the rest frame of hard process is shown for virtuality fixed at 125 GeV,
for up quarks on the left plot and down-quarks on the right plot.
The black solid line is the SM contribution from default implementation in {\tt Tauola++}, and the red dashed line is the SM contribution but using nonSM package.
The effect of nonSM interaction is also shown on the same plots by the blue dotted lines.
\label{PolAng}}
\end{figure}

An  arrangement to verify the proper matching is prepared 
(an  overall $\sqrt{s}$ dependent factors cancel out).
Special printouts from
{\tt TAUOLA/TauSpinner/src/spin2.cxx},
are activated in the DEBUG mode of {\tt TauSpinner}.
Comparison of $\tau$ polarization calculated from {\tt TauSpinner} 
default implementation and the ones from user prepared inputs are then printed.
The corresponding DEBUG mode printout looks as follows:
\begin{verbatim}
test of nonSM Born nonsm2=0
ide,svar,costhe=1 292.05 0.74348
sm=0.529759 sm (new)=0.530185 nsm=0.530185
 sm and smn should be essentially equal.
\end{verbatim}
The {\tt sm} and {\tt smn} denote $\tau$ polarization as obtained from the 
Standard Model respectively by the method of {\tt TauSpinner} and the one of the
user (the third quantity,  {\tt nsm}, provided by the user includes additional 
interaction, which for $\texttt{svar}=s=292.05 ~\text{GeV}^2$ is consistent with zero).

Graphic representation of the above tests are provided in Fig.~\ref{figTest1} as a function of $\sqrt{s}$ for a fixed scattering angle $\theta$,
and in Fig.~\ref{PolAng} for fixed $\sqrt{s}$ and as a function of $\theta$.
In Fig.~\ref{figTest1} consistency up to per mille level on the $\tau$ polarization is obtained
by replacing the {\tt TauSpinner} standard method of Born calculation with the one as defined in Section 2.
For {\tt nonSM2=0} agreement between the user method and {\tt TauSpinner} default should be obtained as in our case.
This is necessary to ensure the non Standard model effect can be correctly implemented with {\tt nonSM2=1}. 
In Fig.~\ref{PolAng}, distributions from all SM contribution as well as new contributions to the $\tau$ polarization are superimposed on the same figure.
Good enough agreement demonstrates the validity of this technical test. 
Further fine tuning of parameters and schemes (e.g. constant or running Z width) in {\tt Tauola++} is not 
necessary\footnote{%
One has to bear in mind that  {\tt TauSpinner} constructs the hard process kinematic configuration differently than {\tt Tauola++}. 
See Ref.~\cite{Czyczula:2012ny} for discussion of potential but minor systematic errors.
}.

Table \ref{Table-pol} represents further tests of $\tau$ polarization, where we check that proper spin states of $\tau$ are provided 
by the {\tt TauSpinner} algorithm using the {\tt getTauSpin()} method (see Ref.~\cite{Czyczula:2012ny} for definition).
Average $\tau$ polarizations are shown when the virtuality of $\tau$ pair is required to lie within a $\pm$ 3 GeV window centered around 125 GeV in the top row.
For the bottom row, a weight = {\tt WT - 1} is used instead of the cut on virtuality. 
Nonetheless, the $\tau$ polarization still includes contributions from SM via the interference effects between the SM and non-SM interactions.
The following cases are monitored:  SM contribution (second column), non-SM contributions for the case of $(L+R)$ couplings as in Section 2 (third column), 
only left coupling for non SM (fourth column) and only  right couplings (fifth column). 
The use of the weight = {\tt WT - 1} is of interest for study of dedicated sub-samples.

\begin{table}[h!]
\begin{center}
{\small
{\begin{tabular}{|c|c|c|c|c|}
\hline
Selection      & $Z\to\tau^+\tau^-$  & $Z/X(L+R)\to\tau^+\tau^-$ & $Z/X(L)\to\tau^+\tau^-$ & $Z/X(R)\to\tau^+\tau^-$ \\
\hline
$125\pm 3$ GeV & -0.448 $\pm$ 0.001  & -0.354 $\pm$ 0.001        & -0.521 $\pm$ 0.001     &   -0.071 $\pm$ 0.001    \\
\hline
{\tt WT - 1}   &  ---                & -0.255 $\pm$ 0.001        & -0.574 $\pm$ 0.001     & \; 0.130 $\pm$ 0.001    \\
\hline
\end{tabular}
}
}
\end{center}
\caption{ 
Average $\tau$ polarization for the SM and non-SM contributions as calculated from helicity states attributed  by standard {\tt TauSpinner} method {\tt getTauSpin().}
} 
\label{Table-pol}
\end{table}

\begin{figure}[htp!]
\begin{tabular}{ccc}
  \includegraphics[width=0.48\columnwidth]{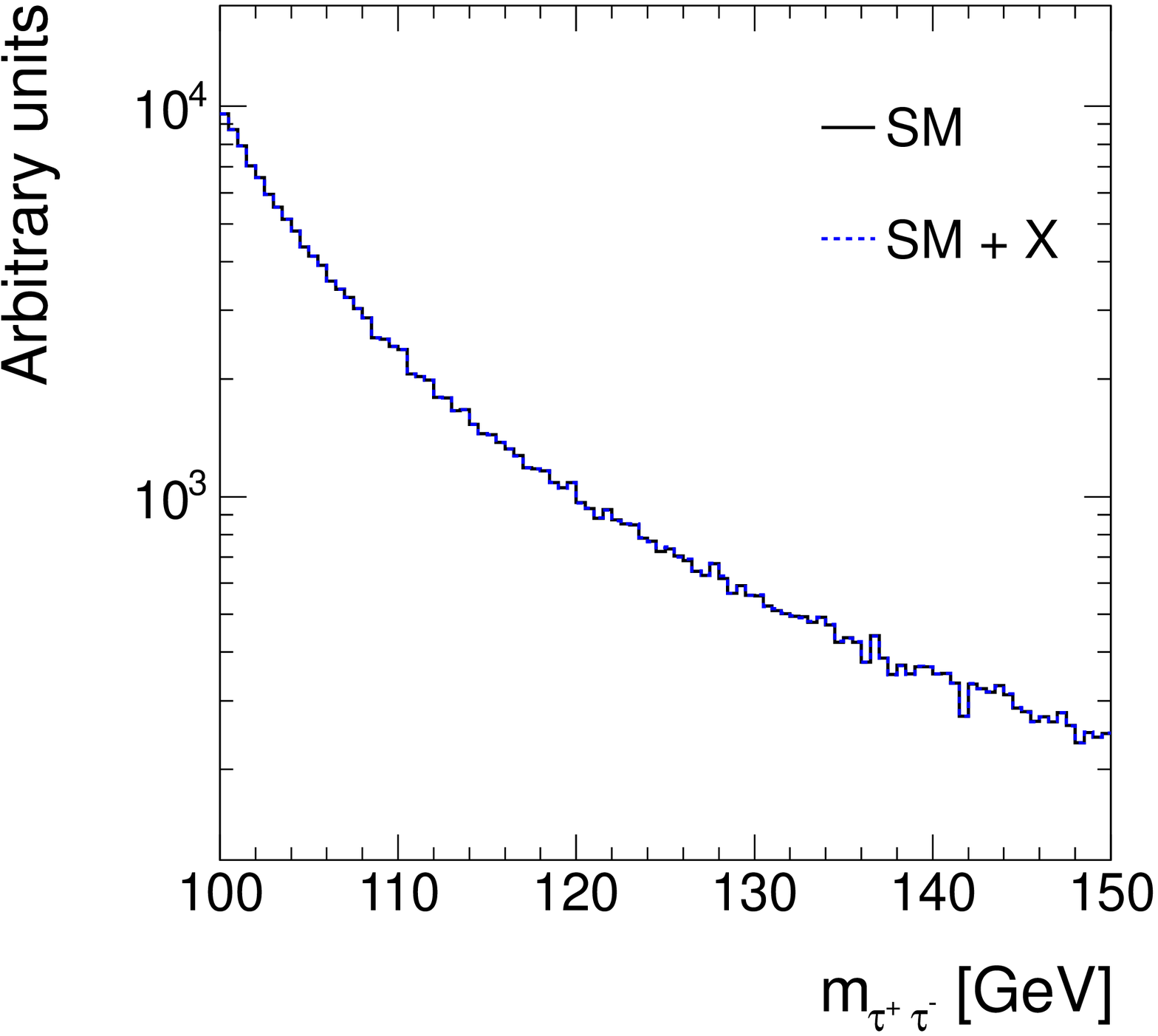} &
  \includegraphics[width=0.48\columnwidth]{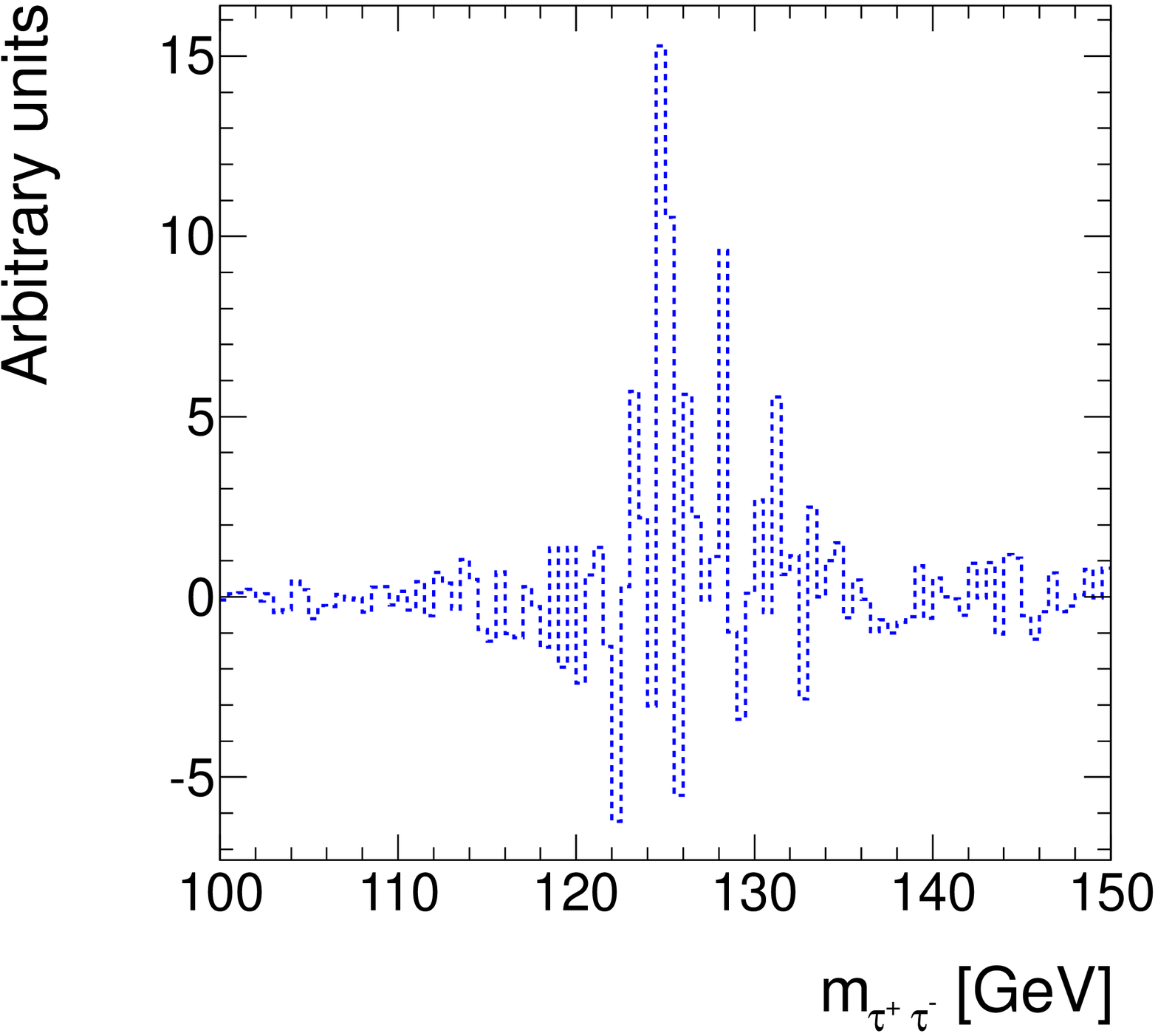} 
\end{tabular}
\caption{Distributions of invariant masses of the $\tau$ pair are shown for the SM contribution in solid black line on the left plot.
The almost overlapping dotted blue line includes the effect from non-SM contribution, 
but  angular integration dependent contributions to the cross section are removed from the weight. 
The right hand side plot visualizes the difference, obtained by applying a weight = {\tt WT - 1}.
\label{figTest2}}
\end{figure}

\begin{figure}[htp!]
\begin{tabular}{ccc}
\includegraphics[width=0.48\columnwidth]{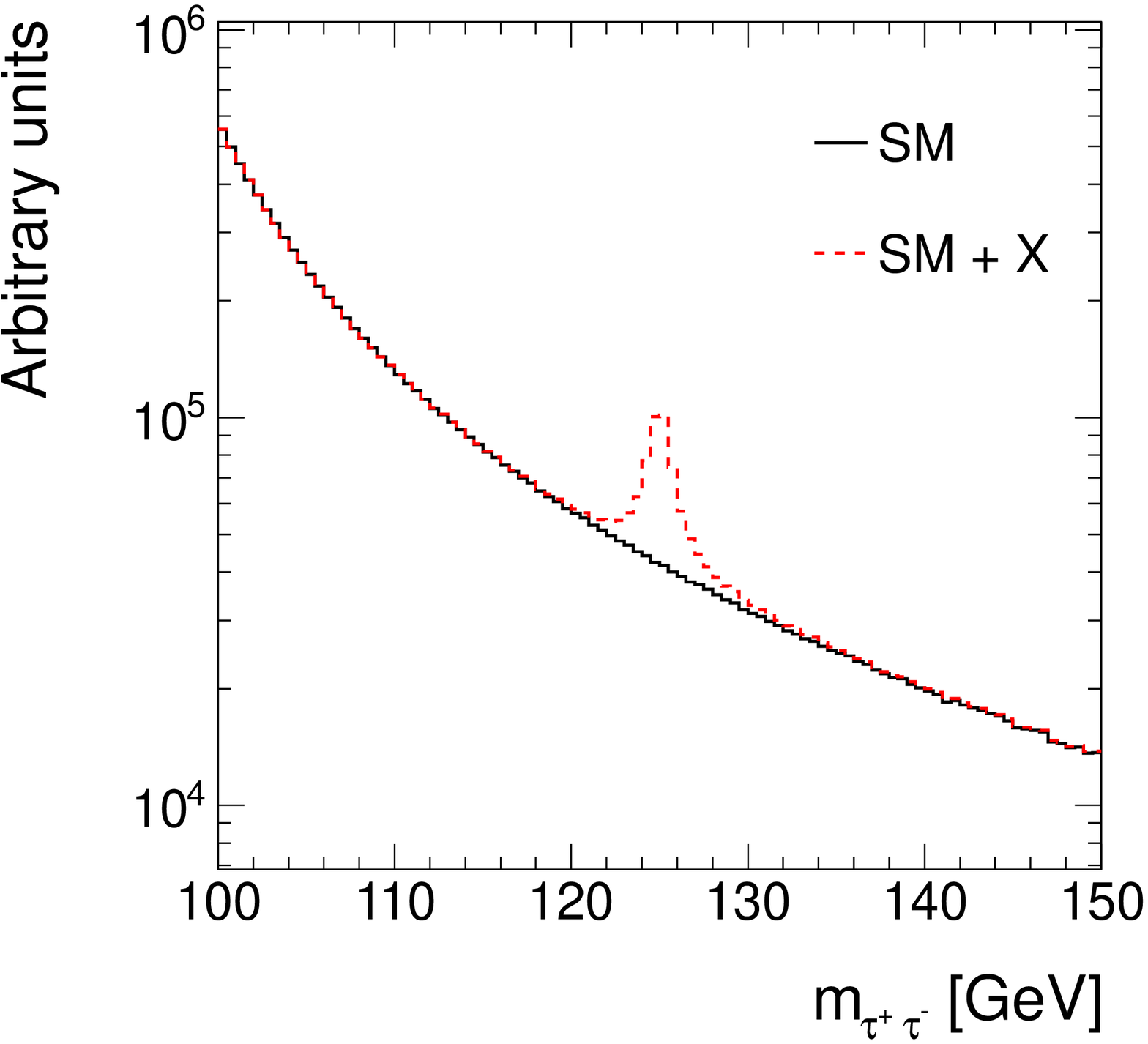} & 
\includegraphics[width=0.48\columnwidth]{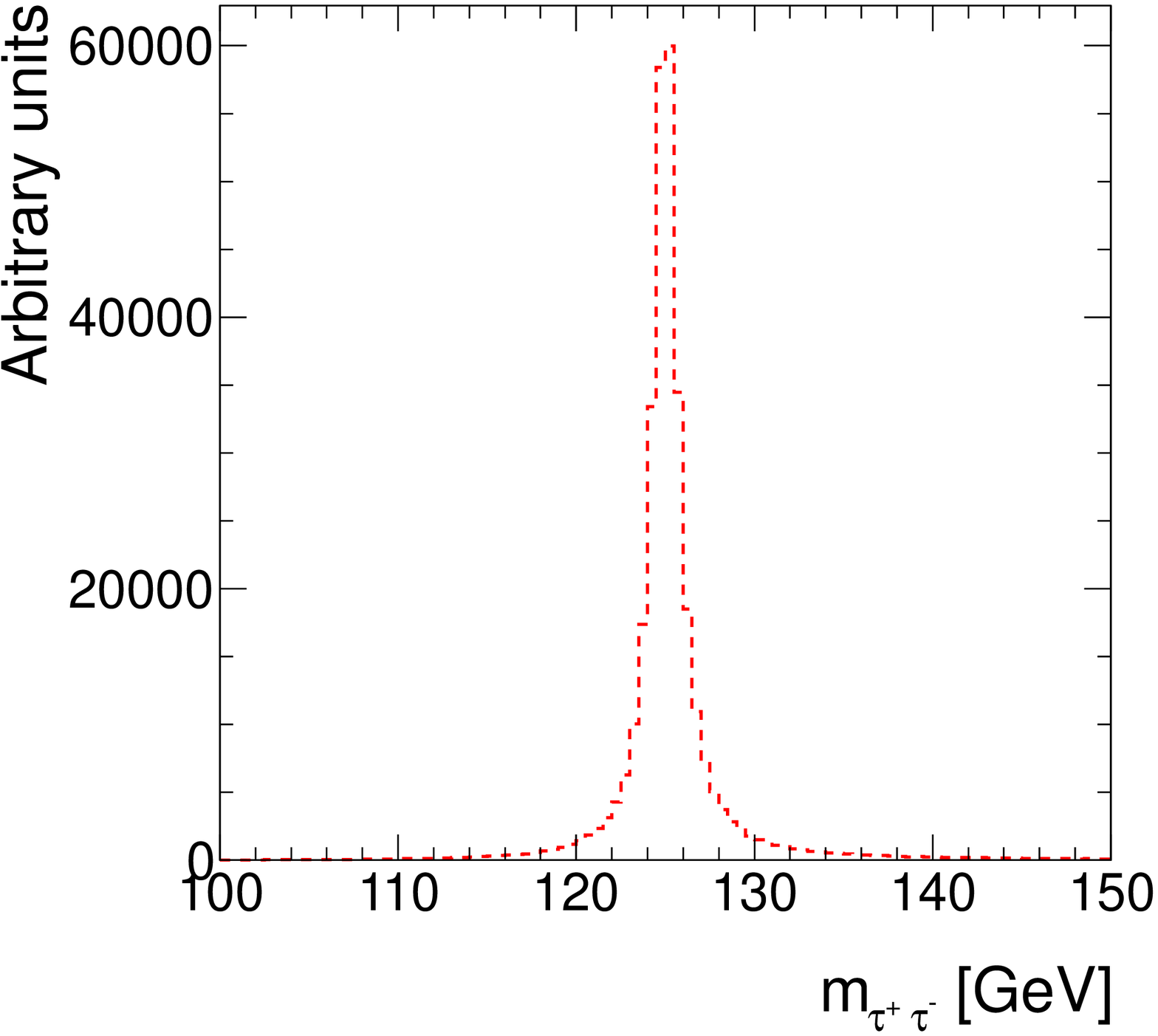} 
\end{tabular}
\caption{Distributions of invariant masses of the $\tau$ pair are shown for the SM contribution in solid black line on the left plot.
The dashed red line includes the effect from non-SM contribution.
The right hand side plot visualizes the difference, obtained by applying a weight = {\tt WT - 1}.
\label{figBase}}
\end{figure}

Second test is performed to check the implementation  of non-SM effects 
when dependence of the cross-section on $\tau$ angular distribution and spin effects are included,
but those arising from integrated Born level cross section are removed. 
In the {\tt TauSpinner} algorithm, we integrate the distribution over $\cos\theta$ 
for the SM cross-sections at the Born level and calculate the effect due to non-SM interactions.
If the calculations are correct, these two contributions from the new matrix element should compensate each other
in the distribution of $\tau$ pair invariant mass $m_{\tau^+\tau^-}$, as visible in Fig.~\ref{figTest2}.
The black solid histogram describing the SM contribution is taken from {\tt Tauola++} by setting both the flags {\tt nonSM2} and {\tt nonSMN} to be equal to 0,
whereas for the blue dotted line the new current from {\tt TauSpinner} is used  but  its effect on cross section are removed. 
It is tested by setting both flags, {\tt nonSM2} and {\tt nonSMN}, to be equal to 1. 
Change of angular distributions resulting from the weight, 
but integrated over $\cos\theta$, results in increased  statistical fluctuations 
as visible on the right hand plot in Fig.~\ref{figTest2} obtained using weight = {\tt WT - 1}.
The amplitude of this procedure is consistent with zero up to per mille level, 
as expected from the size of agreement observed in Fig.~\ref{figTest1}.

The differential effect of not integrating over the angular dependence due to non-SM interactions 
recovers the expected peaking structure around 125 GeV as shown in the red dashed histograms in Fig.~\ref{figBase},
which are obtained by setting the flags {\tt nonSM2} and {\tt nonSMN} to be equal to 1 and 0, respectively.

\section{Spin sensitive observables}

The main feature of {\tt TauSpinner} is that it can work on previously generated and stored data files of the simulated events.
For our purposes, we  study $pp \to Z \to \tau^+ \tau^-$ 
generated at $\sqrt{s}$ = 8 TeV using {\tt Pythia8}~\cite{Sjostrand:2007gs} MC generator, with $\tau$ decays simulated by {\tt Tauola++}~\cite{Davidson:2010rw}.
To apply non-Standard Model weight, 
{\tt TauSpinner} was used\footnote{From {\tt Tauola++} version {\tt v1.1.0}, {\tt TauSpinner} is provided as a part of the {\tt Tauola++} distribution.}.
The distributions for $X\to\tau^+\tau^-$ are obtained by re-weighting the corresponding distributions from $Z\to\tau^+\tau^-$ samples with weight = {\tt WT - 1}.
The samples for gluon-gluon fusion and vector-boson fusion production of the Higgs are generated 
using {\tt POWHEG-BOX}~\cite{Alioli:2008tz,Nason:2009ai} MC generator interfaced to {\tt Pythia8} for showering and hadronization,
while {\tt Pythia8}  is used for the  vector-boson associated Higgs production.
The sample of $H\to\tau^+\tau^-$ events are obtained by summing these three samples 
weighted by their respective cross-sections~\cite{LHCHiggsCrossSectionWorkingGroup:2011ti, LHCHiggsCrossSectionWorkingGroup:2012vm}.

\begin{figure}[htp!]
\begin{tabular}{ccc}
\includegraphics[width=0.48\columnwidth]{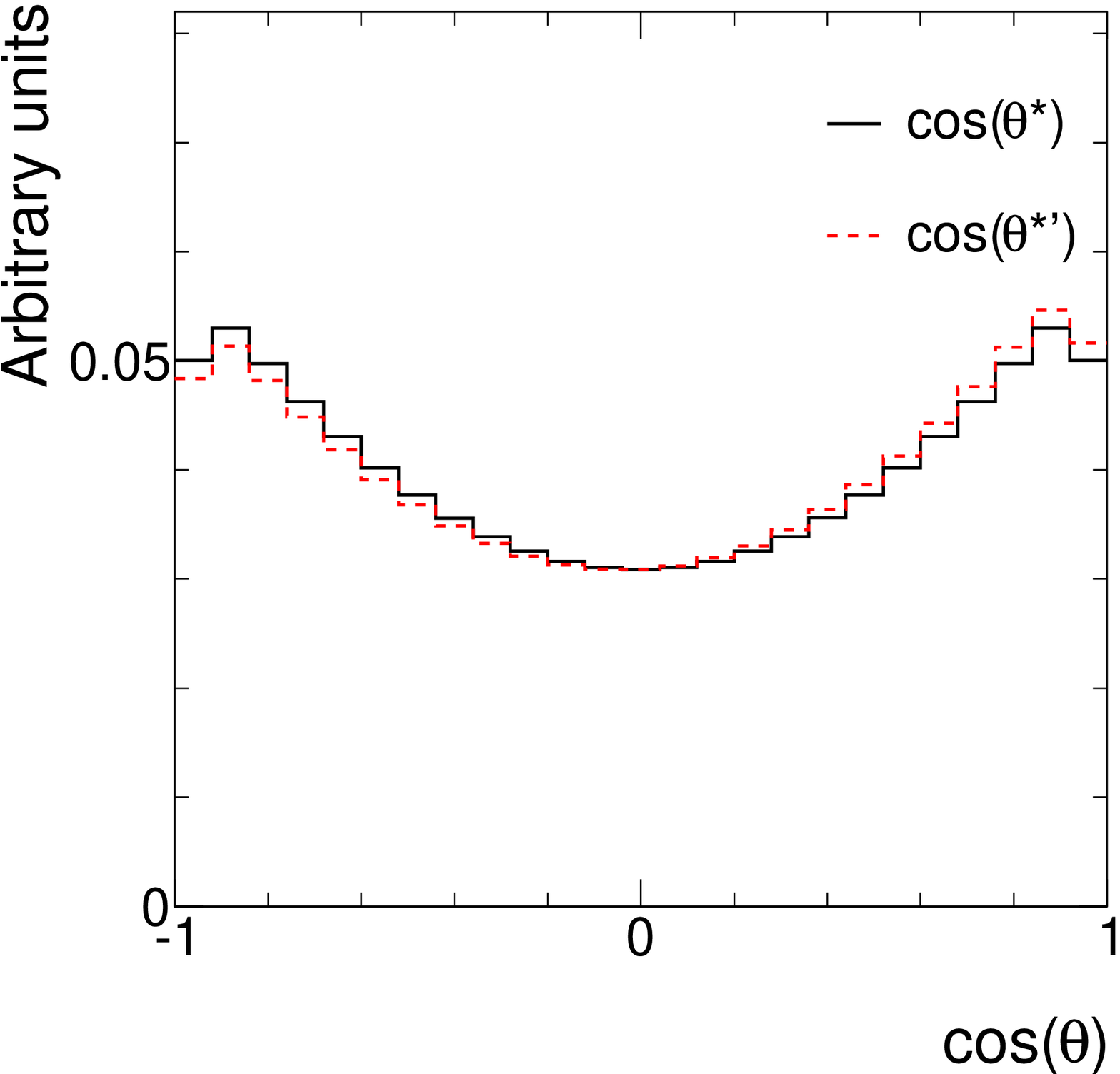} & 
\includegraphics[width=0.48\columnwidth]{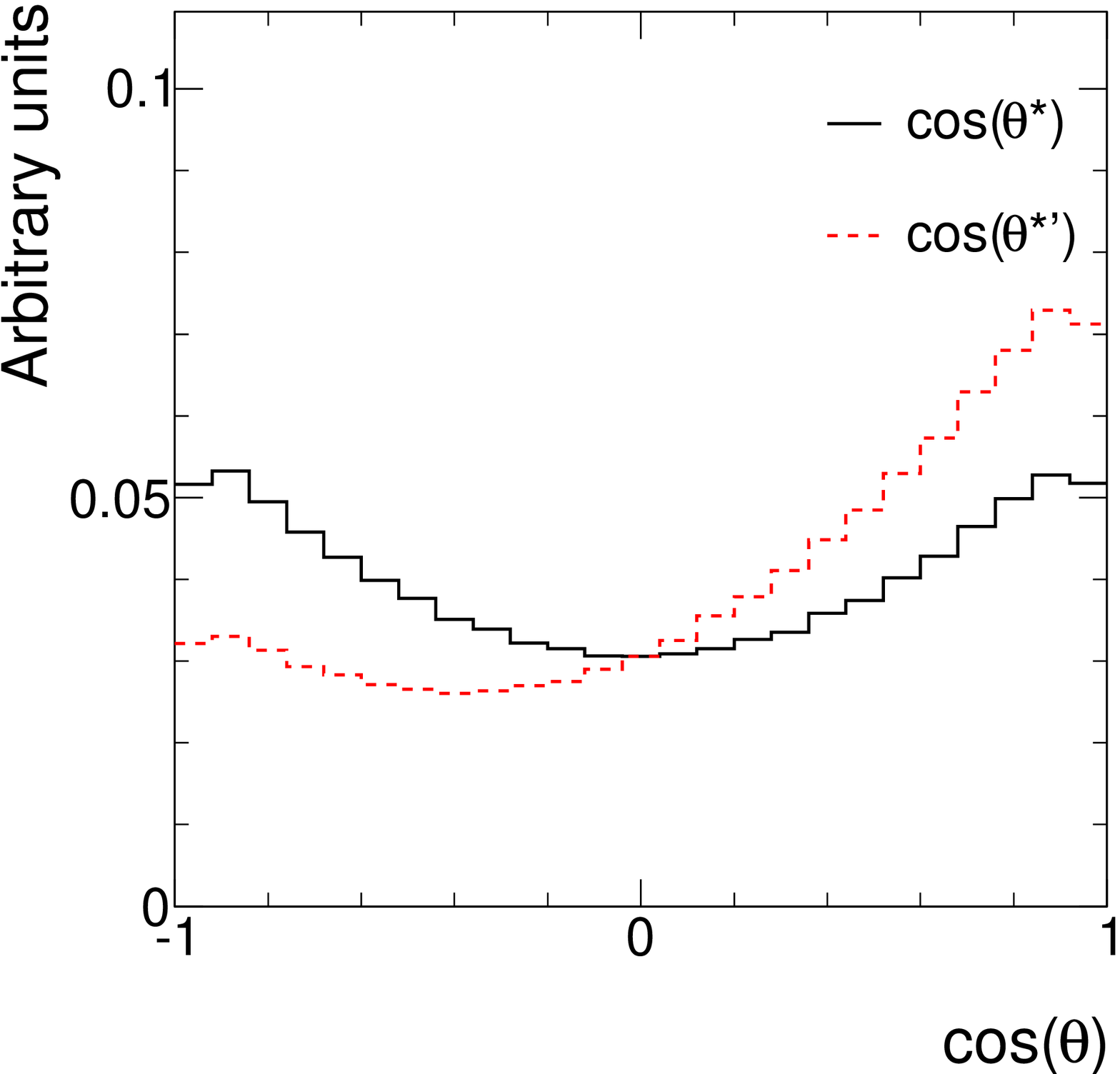} 
\end{tabular}
\caption{The distributions of $\cos(\theta^\star)$ and $\cos(\theta^{\star\prime})$ variables are shown for $Z\to\tau^-\tau^+$ events before any cuts (on the left)
and cuts on truth level invariant mass of the $\tau^+\tau^-$ system restricted to lie within a $\pm$ 3 GeV window centered around 125 GeV (on the right).
All distributions are normalized to unit area.
}
\label{CstarCstarprime_Z}
\end{figure}

\begin{figure}[htp!]
%\begin{center}
\begin{tabular}{ccc}
\includegraphics[width=0.48\columnwidth]{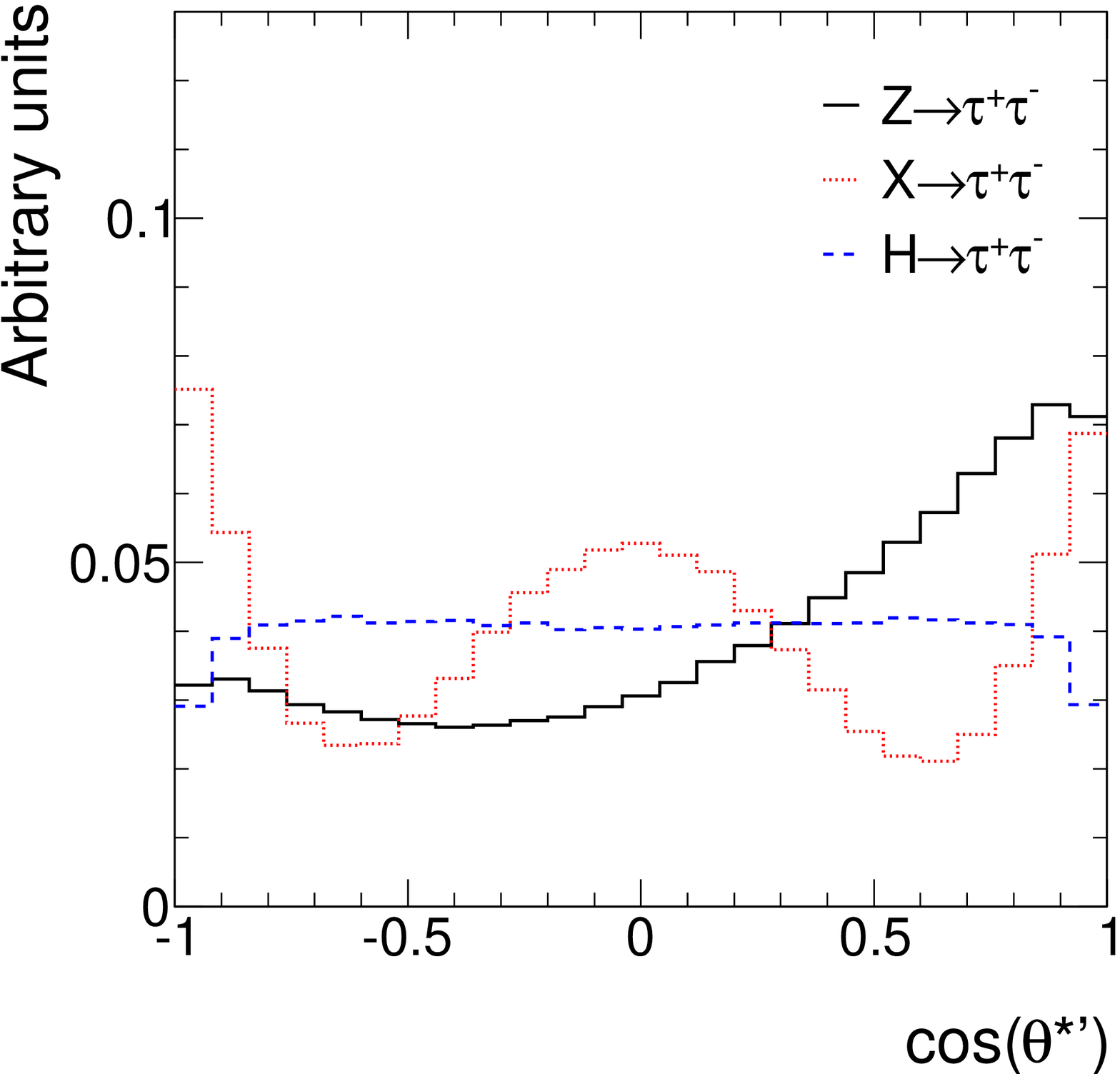} & 
\includegraphics[width=0.48\columnwidth]{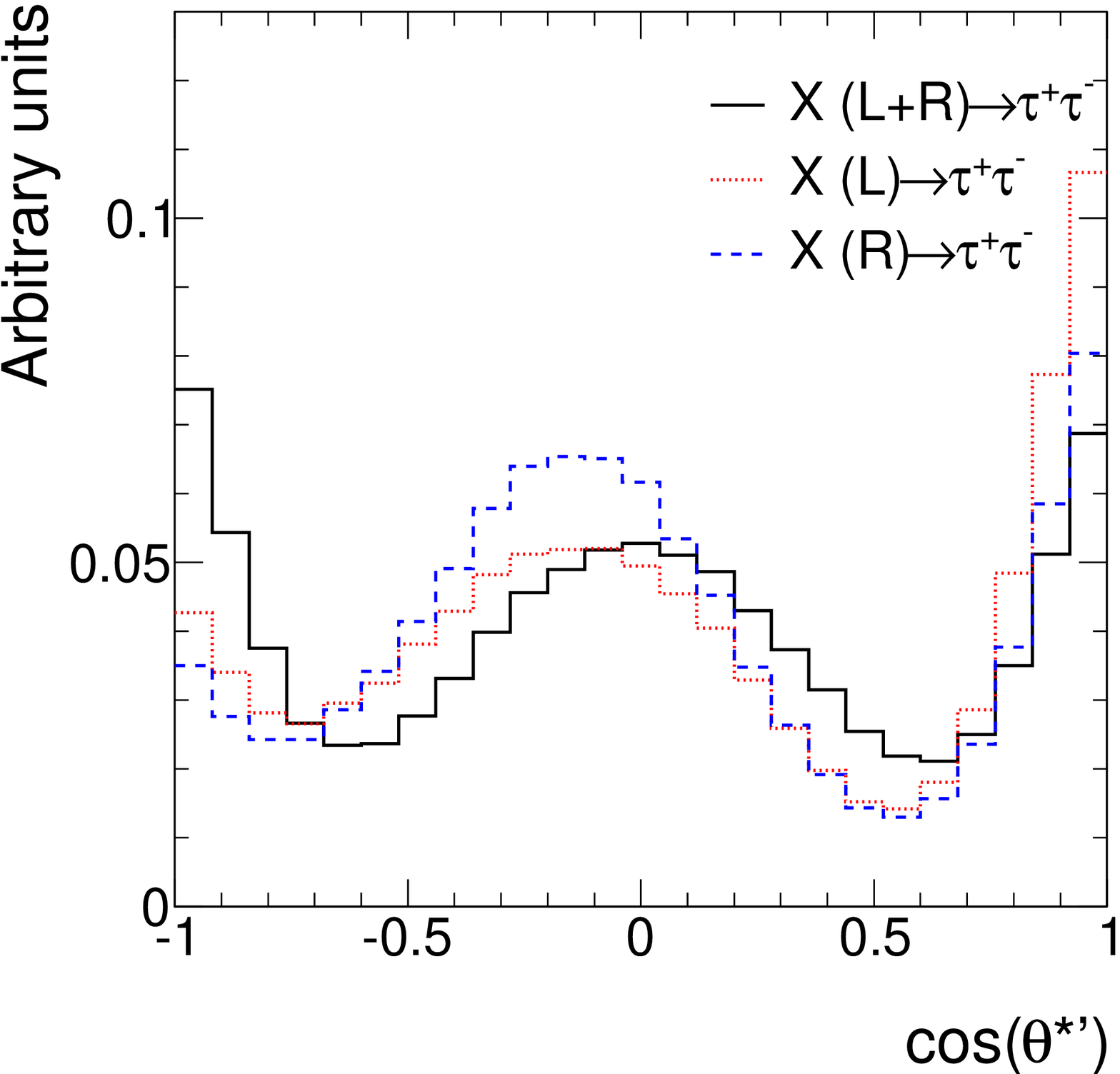} 
\end{tabular}
\caption{The distributions of $\cos(\theta^{\star\prime})$ are shown for $Z\to\tau^+\tau^-$,  $X\to\tau^+\tau^-$, and $H\to\tau^+\tau^-$ events 
after cuts on truth level invariant mass of the $\tau^+\tau^-$ system restricted to lie within a $\pm$ 3 GeV window centered around 125 GeV (on the left).
For the same selection criteria, the distributions of $\cos(\theta^{\star\prime})$ are shown (on the right) 
for $X\to\tau^-\tau^+$ events for three choices of coupling constants as described in the text.
All distributions are normalized to unit area.
}
\label{Cstarprime_ZXH}
\end{figure}

We study the variable $\cos(\theta^\star)$~\cite{Was:1989ce},
describing the average scattering angle between the observable final state products from $\tau^+$ and $\tau^-$ decays, respectively.
In the leading approximation, it corresponds to scattering angle in the rest-frame of hard process.
In the laboratory frame, this variable is defined as:
\begin{equation}
\cos(\theta^\star)= \frac{\sin\theta^{\tau^-}\cos\theta^{\tau^+} + \sin\theta^{\tau^+}\cos\theta^{\tau^-} }{\sin\theta^{\tau^+}+ \sin\theta^{\tau^-} },
\end{equation}
where $\theta^{\tau^+}$ is the angle between the decay products from $\tau^+$ and negative z-axis,
and $\theta^{\tau^-}$ is the angle between the  decay products from $\tau^-$ and positive z-axis.

Sensitivity to forward-backward asymmetry is enhanced by flipping the sign of the variable $\cos(\theta^\star)$,
whenever the vector sum of momentum of the visible $\tau$ daughters is aligned towards the negative z-axis.
This defines the variable $\cos(\theta^{\star\prime})$, which is sensitive to the spin of Z/X/H.

The distributions of $\cos(\theta^\star)$ and $\cos(\theta^{\star\prime})$ are shown in Fig.~\ref{CstarCstarprime_Z} for $Z\to\tau^+\tau^-$ events.
The left plot in Fig.~\ref{CstarCstarprime_Z} compares the distributions before any cuts.
The right plot compares the distributions after a cut on virtuality has been applied 
by selecting events with the truth level invariant mass of the $\tau^+\tau^-$ system restricted to lie within a $\pm$ 3 GeV window centered around 125 GeV.

The distributions of $\cos(\theta^{\star\prime})$ for $X\to\tau^+\tau^-$, and $H\to\tau^+\tau^-$ events are shown on the left plot in Fig.~\ref{Cstarprime_ZXH}.
The effect of spin is clearly visible in this variable which shows striking difference as compared to $Z\to\tau^+\tau^-$ events, also shown in the same plot.
The virtuality of all these processes are chosen to lie within a $\pm$ 3 GeV window centered around 125 GeV 
by applying a cut on the truth level invariant mass of the $\tau^+\tau^-$ system.

For the same selection criteria, the distribution of $X\to\tau^-\tau^+$ events are shown 
on the right plot in Fig.~\ref{Cstarprime_ZXH} for different choices of coupling constants.
As expected from Table~\ref{Table-pol}, the asymmetry in this variable is strongest for the case of pure left-handed coupling,
and weakest for the case of pure right-handed coupling. 

\section{Test for spin of $Z/X/H$}

A key feature of  $\tau$ polarization in $Z\to\tau^+\tau^-$ events is its dependence on virtuality of the hard interaction.
To select the $Z\to\tau^+\tau^-$ events as relevant backgrounds for a Higgs-like particle with a mass of 125 GeV,
we select events inside an appropriate window on the collinear mass ($m^{\rm{coll}}_{\tau^+\tau^-}$) reconstructed from the $\tau$ decay products~\cite{Ellis:1987xu}.
This is sufficient for our illustrative purposes, and is comparable to computationally intensive techniques developed for di-tau mass reconstruction~\cite{Elagin:2010aw}.

\begin{figure}[htp!]
\begin{tabular}{ccc}
\includegraphics[width=0.48\columnwidth]{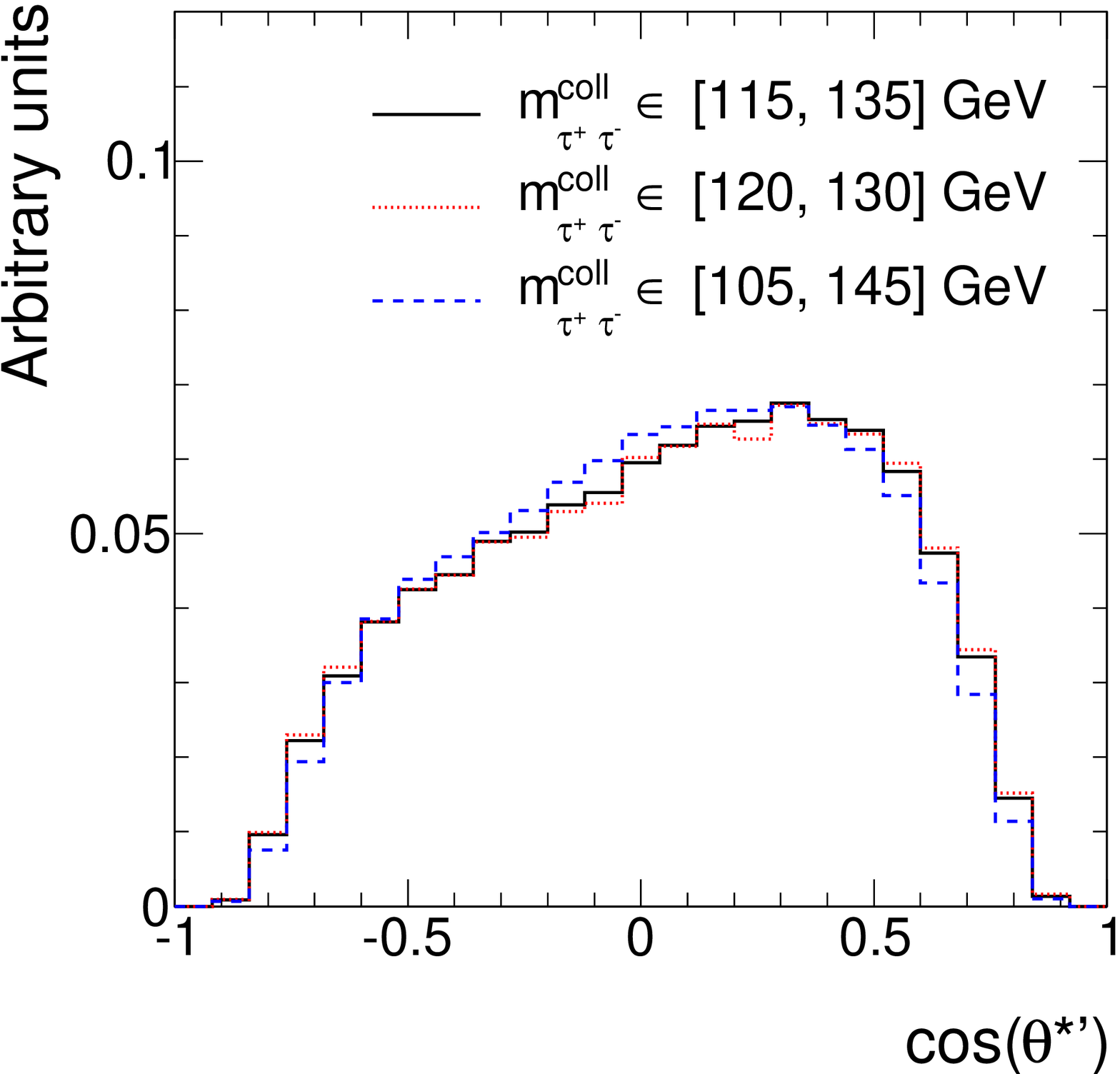} & 
\includegraphics[width=0.48\columnwidth]{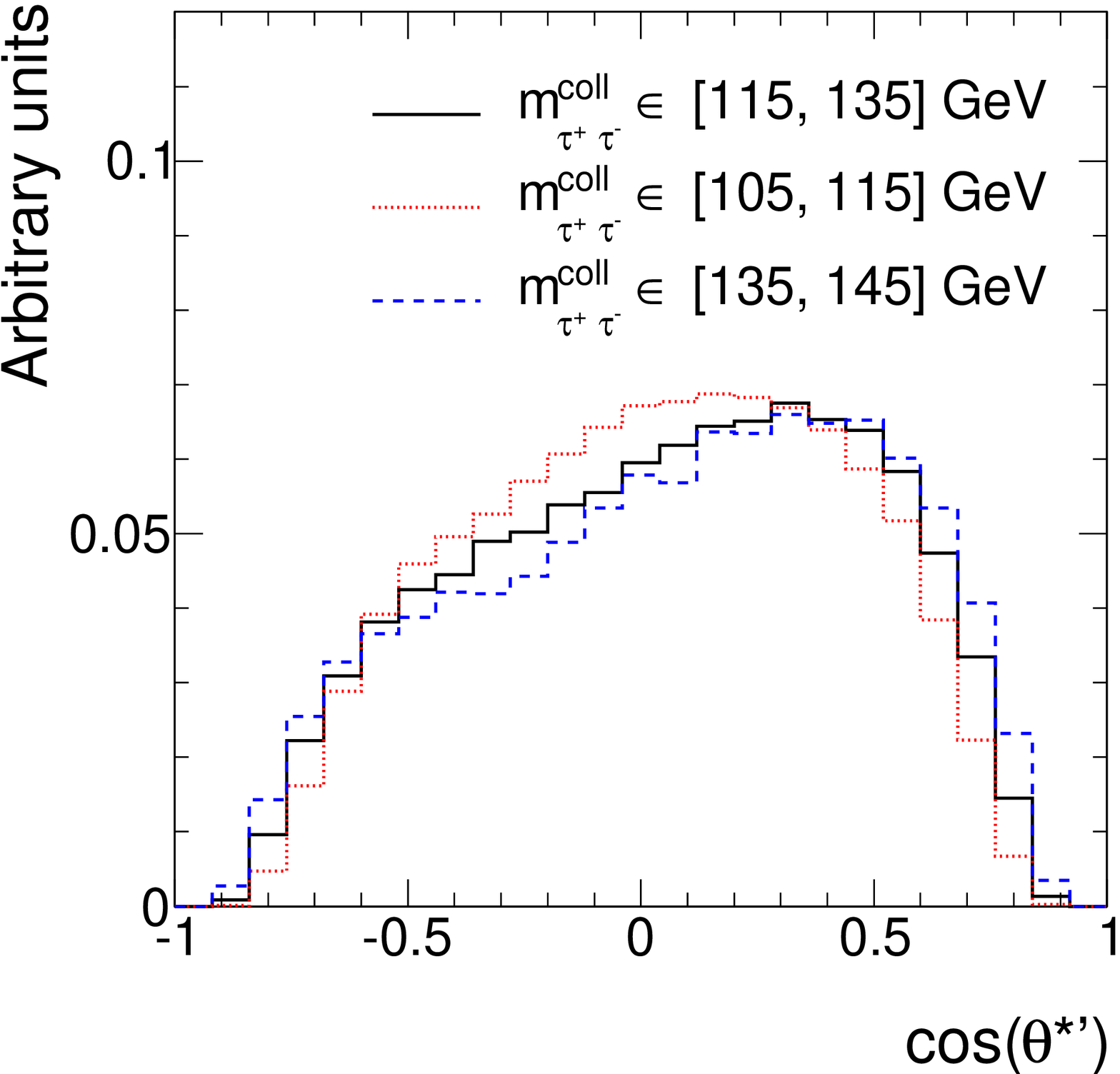} 
\end{tabular}
\caption{The distributions of $\cos(\theta^{\star\prime})$ variable are shown for $Z\to\tau^+\tau^-$ events with cuts on reconstructed level quantity $m^{\rm{coll}}_{\tau^+\tau^-}$.
The plot on the left corresponds to $\pm$ 10, $\pm$ 5 and $\pm$ 20 GeV windows centered around 125 GeV.
The plot on the right corresponds to $\pm$ 10, $\pm$ 5 and $\pm$ 5 GeV windows centered around 125, 110 and 140 GeV, respectively.
All distributions are normalized to unit area.
}
\label{Cstarprime_cuts_Z}
\end{figure}

\begin{figure}[htp!]
\begin{tabular}{ccc}
\includegraphics[width=0.48\columnwidth]{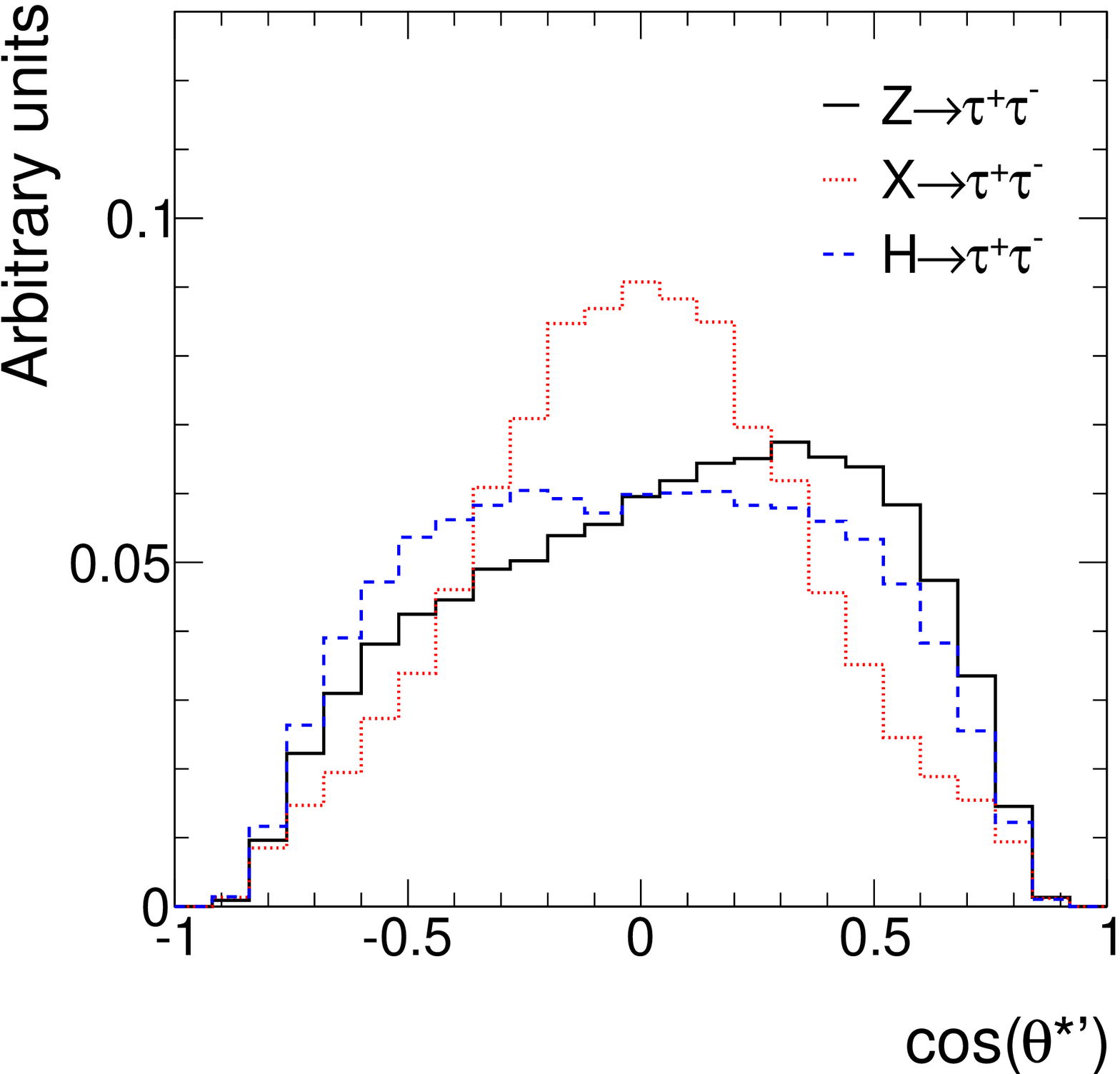} & 
\includegraphics[width=0.48\columnwidth]{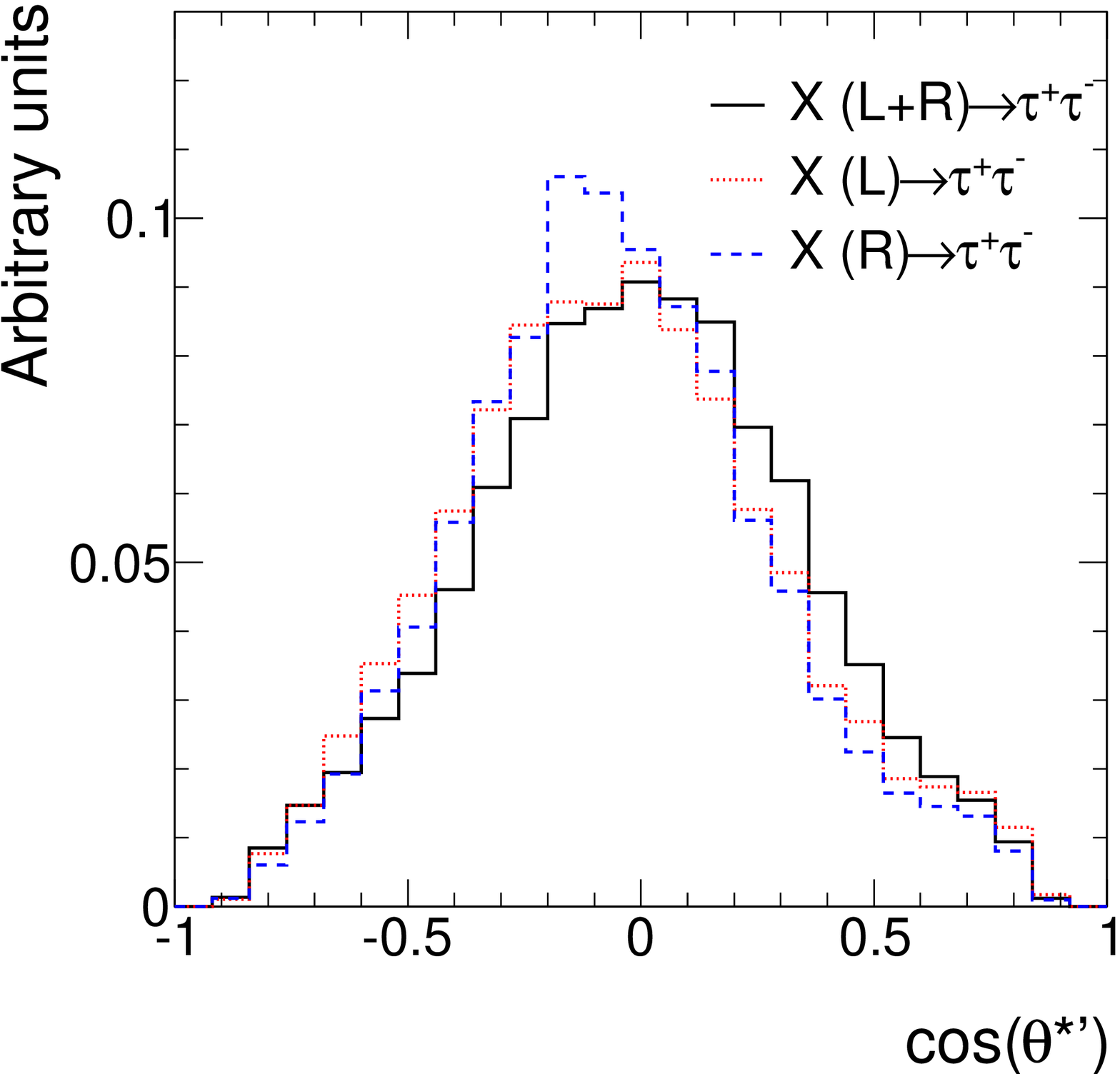} 
\end{tabular}
\caption{
The distributions of $\cos(\theta^{\star\prime})$ are shown (on the left) for $Z\to\tau^+\tau^-$, 
$X\to\tau^+\tau^-$, and $H\to\tau^+\tau^-$  events after cuts as described in the text.
The distributions of $\cos(\theta^{\star\prime})$ are shown (on the right) after cuts for $X\to\tau^+\tau^-$ events 
with different options for coupling constants as described in the text.
All distributions are normalized to unit area.
}
\label{Cstarprime_cuts}
\end{figure}

Our selection criteria are similar to the ones as in Ref.~\cite{Rainwater:1998kj, Plehn:1999xi}, 
and those used by the ATLAS~\cite{ATLAS_HCP} and CMS~\cite{CMS_HCP} collaborations.
The following requirements are applied to select sample of events enriched with a new resonance produced at 125 GeV:
\begin{itemize}
\item $p_{\rm{T}}$ for each of the visible daughters of the tau's are  required to be greater than 20 GeV and lie within an acceptance of $|\eta|<2.5$,
\item the missing transverse momentum ($E_{\mathrm{T}}^{\mathrm{miss}}$), 
defined as the $p_T$ of the vector sum of the neutrino's momentum, is required to be greater than  20 GeV,
\item the transverse mass of the system comprising of $E_{\mathrm{T}}^{\mathrm{miss}}$ and the visible daughter with smaller $p_{\rm{T}}$ 
defined as $\sqrt{2p_{\rm{T}}E_{\mathrm{T}}^{\mathrm{miss}}(1-\cos\Delta\phi)}$ is required to be less than 50 GeV, 
where $\Delta\phi$ is the angle between the directions of $E_{\mathrm{T}}^{\mathrm{miss}}$ and visible daughter 
with smaller $p_{\rm{T}}$ in the plane perpendicular to the beam direction,
\item the cosine of the 3-dimensional opening angle between the two  daughters is required to be greater than -0.9,
\item the difference in azimuthal angles between the two daughters is required to be less than 2.9, and
\item $m^{\rm{coll}}_{\tau^+\tau^-}$ is required to lie within a $\pm$ 10 GeV window centered around 125 GeV.
\end{itemize}

The stability of the distribution of the discriminating variable $\cos(\theta^{\star^\prime})$ for $Z\to\tau^+\tau^-$ events 
with respect to selected region of virtuality is studied in Fig.~\ref{Cstarprime_cuts_Z}.
The requirements on the window on  $m^{\rm{coll}}_{\tau^+\tau^-}$ is varied, keeping all the other above-mentioned selection requirements the same.
The left plot corresponds to $\pm$ 10, $\pm$ 5 and $\pm$ 20 GeV windows centered around 125 GeV. 
The right plot corresponds to $\pm$ 10, $\pm$ 5 and $\pm$ 10 GeV windows centered around 125, 110 and 140 GeV, respectively.
For the case when virtuality is centered around 125 GeV, the distributions are compatible, while the shape is altered with shift of the center of the window.
Thus, these differences in shapes can be taken as estimates of systematic uncertainties on the modeling of the background from $Z\to\tau^-\tau^+$ events.

The left plot in Fig.~\ref{Cstarprime_cuts} shows the distribution of $\cos(\theta^{\star\prime})$ after applying these selection requirements
to sample of $Z\to\tau^+\tau^-$,  $X\to\tau^+\tau^-$, and $H\to\tau^+\tau^-$ events. 
For illustrative purposes, the distributions of $\cos(\theta^{\star\prime})$ are shown in the right plot of Fig.~\ref{Cstarprime_cuts} 
for $X\to\tau^+\tau^-$ events passing these selection criteria corresponding to the three different choices of the coupling constants.

Visible separation between the shapes of $H\to\tau^+\tau^-$ and $X\to\tau^+\tau^-$ events are observed, 
which is similar in size for all the three choices of coupling constants studied.
The distributions from  $H\to\tau^+\tau^-$ events are incompatible at the 99\% confidence level with those arising from $X\to\tau^+\tau^-$ events
for a sample of 500 Higgs-like events, which corresponds to roughly twice the number of signal events 
observed with 13 ${\rm{fb}}^{-1}$ at $pp$ collision energy of 8 TeV by the ATLAS collaboration~\cite{ATLAS_HCP}.
However, the separation power depends on the presence of backgrounds and choice of selection criteria, 
which alters relative efficiencies of signal and background events.
Further discrimination power can be obtained by categorizing the $\tau^+\tau^-$ decay modes 
depending upon the observed final state particles.

\section{Summary}

In this note we have presented a new development of the {\tt  TauSpinner} 
that is capable of addressing possible extension of the SM based on new physics model 
that results in contribution from a spin 2 state $X$. 
To simplify estimation of observability of such a state at the LHC,
extension for the {\tt TauSpinner} algorithm to manipulate $\tau$ pair final states 
in previously generated MC samples is proposed  which employs the method of event weights.
Weights calculated with {\tt TauSpinner}  feature an implementation of amplitudes from beyond the Standard Model.
Not only spin correlation effects as in the previous versions, 
but also effects modifying the angular distributions of $\tau^\pm$ lepton directions 
and size of the cross-section can be studied in this way.

An  example of the installation of our new  matrix elements into the  {\tt TauSpinner} algorithm is presented.
Distributions validating  correctness of the installation are discussed and  are found correct.

We then study distributions of experimentally observable quantities sensitive to the spin of $Z/X/H$.
In semi-realistic conditions, we study the impact of the new interaction and its signature. 
We found the approximate hard process scattering angle $\theta^{\star\prime}$ 
reconstructed from observable directions of $\tau$ decay products useful for that purpose.
Following the experimental searches performed with half the dataset expected to be collected at $\sqrt{s}$ = 8 TeV in 2012,
we apply selection criteria that enrich the sample of events with a Higgs-like state with mass $\sim$ 125 GeV.
With double the number of Higgs-like events as can be expected using full dataset in 2012, 
we find distributions that are incompatible between $H\to\tau^+\tau^-$ and $X\to\tau^+\tau^-$ events at the 99\% confidence level.
Detailed experimental study with all background contributions, which include detector resolution and acceptance effects, 
as categorized over the different $\tau^+\tau^-$ decay modes is expected to improve the discrimination power.

Robustness of the method was demonstrated here, and the first results are of potential interest. 
Method is also straightforward to extend to other cases such as $Z'$ etc. 
Alternative production mechanism of spin-2 states, eg. via gluon-gluon fusion, 
may also be implemented by appropriate re-weighting of the angular dependence from corresponding matrix elements.
A key aspect of our  {\tt TauSpinner} algorithm is that computationally expensive simulation 
of independent MC samples is not necessary for study of new physics interactions.

\appendix
\section{ {\tt TauSpinner} - changes introduced in version 1.2}
\label{sec:TauSpinnerChanges}
Since its first public version, described in \cite{Czyczula:2012ny}, two
new updates to {\tt TauSpinner} has been introduced. First, {\tt TauSpinner}
has been merged into {\tt Tauola++}\cite{Davidson:2010rw} distribution and now,
while working on this paper, it has been extended to add new functionality.
In this section we  list the changes between version 1.0 and 1.2.

\begin{itemize}
\item Merging with {\tt Tauola++} \\
      {\tt TauSpinner} now comes as an additional library to {\tt Tauola++}
      distribution. {\tt Tauola++} configuration scripts have been updated
      to accomodate this setup. As of writing this paper, {\tt Tauola++ v1.1.1},
      featuring {\tt TauSpinner v1.2} has been installed in
      {\tt GENSER}\cite{LCG,Kirsanov:2008zz} database.
\item Two new initialization options - {\tt nonSM2} and {\tt nonSMN} \\
      The {\tt nonSM2} flag turns on non-Standard Model weight calculation.
      The {\tt nonSMN} flag, combined with {\tt nonSM2}, allows for calculation
      of corrections to shape only.
\item New functions added. \\
      An example {\tt examples/tau-reweight-test.cxx} has been updated to show
      how functions described below can be used in case of spin-2 calculation
       described in this paper.
      \begin{itemize}
      \item {\tt set\_nonSM\_born( double (*fun)(int ID, double S, double cost, int H1, int H2, int key) ) } \\
            Sets function for user-defined born, including new physics.
            The parameters of this new function are described in {\tt include/TauSpinner/nonSM.h}
            as well as in the example program.
      \item {\tt void setNonSMkey(int key)} \\
            Sets the value of {\tt nonSM2} flag. Allows turning non-Standard Model
            calculation on and off for comparison between different models.
      \item {\tt double getWtNonSM()} \\
            Returns non-Standard Model weight {\tt WT3} calculated for the last event
            processed by {\tt TauSpinner}.
      \end{itemize}
\end{itemize}

In our example {\tt examples/tau-reweight-test.cxx} fortran function is provided to calculate 
quark level Born cross section where new physics effects can be switched on and off.
Physics model described in the previous sections
at the level of quark level annihilation into pair of tau leptons is used. Function: \\
\\
{\tt REAL*8 FUNCTION DISTJWK(ID,S,T,H1,H2,KEY)} \\
\\
is used in our program with the help of the C++ function: \\
\\
{\tt  double nonSM{\_}adopt(int ID, double S, double cost, int H1, int H2, int key)} \\
\\
Its use is initialized with the method 
{\tt set{\_}nonSM{\_}born( nonSM{\_}adopt ); }

Other user defined function can be used in the same way.

\section*{Acknowledgements}
Encouragement, support and discussions with Sau Lan Wu are acknowledged.
Useful discussions with Ian Nugent are acknowledged.
This project is financed in part from funds of Polish National Science
Centre under decision DEC-2011/03/B/ST2/00107 (TP), DEC-2011/03/B/ST2/00220 
(ZW) and DEC-2011/01/M/ST2/02466 (JK). 

\bibliography{Tauola_interface_design}{}
\bibliographystyle{utphys_spires}

\end{document}